\documentclass[aps,prl,twocolumn,superscriptaddress,10pt]{revtex4}

\usepackage{graphicx}
\usepackage{amsmath,amssymb,amstext,amsthm}
\usepackage{bbold}
\usepackage{cancel}
\usepackage{color}
\usepackage{verbatim}
\usepackage{booktabs}
\usepackage[caption=false]{subfig}
\usepackage[english]{babel}

\begin{document}

\title{Higher-order topological insulators protected by inversion and rotoinversion symmetries}

\author{Guido van Miert}
\affiliation{Institute for Theoretical Physics, Center for Extreme Matter and
Emergent Phenomena, Utrecht University, Princetonplein 5, 3584 CC Utrecht,
Netherlands}

\author{Carmine Ortix}
\affiliation{Institute for Theoretical Physics, Center for Extreme Matter and
Emergent Phenomena, Utrecht University, Princetonplein 5, 3584 CC Utrecht,
Netherlands}
\affiliation{Dipartimento di Fisica ``E. R. Caianiello", Universit\'a di Salerno, IT-84084 Fisciano, Italy}

\date{\today}

\begin{abstract}
We provide the bulk topological invariant for chiral higher-order topological insulators in: {\it i}) fourfold rotoinversion invariant bulk crystals, and {\it ii}) inversion-symmetric systems with or without an additional three-fold rotation symmetry. 
These states of matter are characterized by a non-trivial $\mathbb{Z}_2$ 
index,
which we define in terms of symmetric hybrid Wannier functions of the filled bands, and can be readily calculated from the knowledge of the crystalline symmetry labels of the bulk band structure. The topological invariant determines the generic presence or absence of protected chiral gapless one-dimensional modes localized at the hinges between conventional gapped surfaces.  
\end{abstract}

\maketitle

\paragraph{Introduction --} 
Free electrons in a crystal can be universally described in terms of Bloch waves.  In ordinary band insulators, however, electronic states can be equivalently represented using exponentially localized Wannier functions (WFs), {\it i.e.} the Fourier transform of the Bloch waves up to a unitary basis transformation. When (non)-spatial symmetries are involved, this gauge degree of freedom can be exploited to construct WFs respecting the set of symmetries characterizing the insulating crystal~\cite{mar97,mar12,sak13}. 
Symmetry-protected topological (SPT) insulators do not allow for such a Wannier representation. 
This is because the most generic SPT cannot be adiabatically deformed to a trivial atomic insulator~\cite{bra17,po17}, whose orbitals naturally form a set of symmetric WFs. Henceforth, the presence of ``anomalous" gapless boundary states~\cite{has10,qi11} -- the prime physical consequence of a bulk non-trivial topology -- can be related to an obstruction in representing the ground state of the system in terms of symmetric WFs~\cite{sol11,po17}. 

Nevertheless, a description of a SPT state in terms of symmetric hybrid Wannier functions (HWFs), the partial Fourier transform of Bloch waves,  is entirely allowed~\cite{coh09}. An inversion symmetric Chern insulator, for instance, can be described using HWFs obeying the property $w^{\textrm{hybrid}}_k(x) \equiv w^{\textrm{hybrid}}_{-k}(-x)$. 
At the two momenta $k=0,\pi$ the HWFs thus correspond to inversion-symmetric one-dimensional (1D) WFs whose charge centers constitute genuine topological invariant~\cite{hug11}, and dictate the existence of quantized end-charges in open geometries~\cite{mie16,rhi17,mie17}. 
These effective 1D crystalline topologies are not only interesting {\it per se}. In fact, and as shown below, their mismatch allows, in a very simple manner, to diagnose the parity of the integer Chern number characterizing the two-dimensional topological insulating state. 

Starting out from this observation, in this 
Rapid Communication we
define the bulk topology of second-order topological insulators~\cite{sit12,lan17,son17,eza18,sch18,gei18,kha18,fan17,sch18bis,Eza18a,Eza18b} --  crystalline systems with gapped $(d-1)$ surface states but gapless $(d-2)$ chiral hinge modes -- protected 
only
by spatial (roto)inversion symmetries. To this end, we will first define the crystalline topology of rotationally-symmetric two-dimensional (2D) insulators with zero Chern number using symmetric WFs. The corresponding topological invariants, which can be also expressed in terms of the symmetry labels in the band structure representation~\cite{ben14,kru17}, directly pinpoint the presence of quantized corner charges in open geometries~\cite{zho15,ben17,ben17bis,xu17,ser18,pet18,imh17}.  Thereafter, we will show that a mismatch between these 2D crystalline topological invariants in the symmetric HWFs representation of a bulk 
three-dimensional
crystal, encoded in a $\mathbb{Z}_2$ topological invariant, dictates the existence of protected chiral hinge modes. In the remainder, we present explicit calculations for fourfold rotoinversion symmetric crystals and refer to the Supplemental Material~\cite{notesup} for other point-group symmetries. 

\begin{figure}[tbp]\centering
\includegraphics[width=.85\columnwidth]{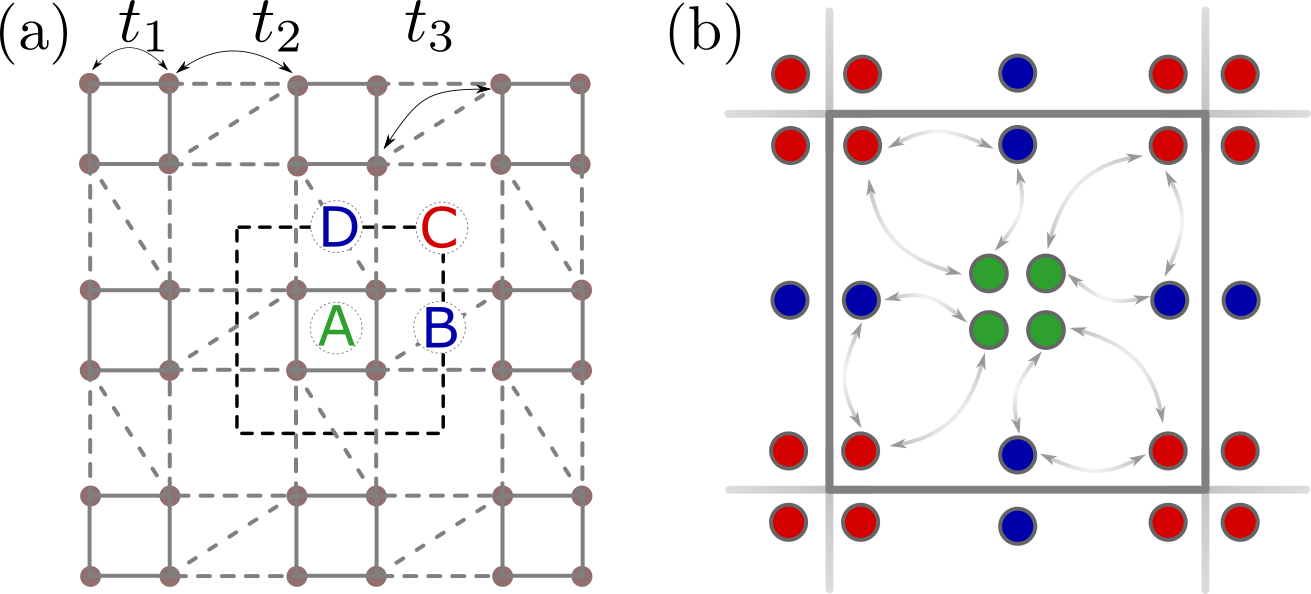}
\caption{(Color online) (a) A tight-binding model for a $\mathcal{C}_4$-symmetric crystal with four atoms in the unit cell. We explicitly indicate the hopping amplitudes and the Wyckoff positions in the unit cell. (b) Three topologically equivalent orbital configurations with different sets of $N_{W;r}$ integers.}
\label{fig:WannierC4}
\end{figure}

\paragraph{Crystalline topology of 2D insulators via symmetric WFs. -- } 
We first recall the concept of Wyckoff positions. General positions in the unit cell can be classified into a few types of Wyckoff positions depending upon their site symmetry group. In particular, for a two-dimensional crystal with $\mathcal{C}_4$ rotational symmetry, there are two Wyckoff positions that are invariant under fourfold rotation [c.f. Fig.~\ref{fig:WannierC4}(a)], which can be fixed at the origin ${\bf r}=\left\{0,0 \right\}$ ($A$) and at the corner of the unit cell ${\bf r}=\left\{1/2, 1/2 \right\}$ ($C$). The two sites ${\bf r}=\left\{1/2 , 0\right\}$ ($B$), and ${\bf r}=\left\{0, 1/2 \right\}$ ($D$) are instead separately invariant under twofold rotations, but are transformed into each other by a fourfold rotation. 
This also implies that symmetric WFs centered at the Wyckoff positions $A,C$ are classified~\cite{notesup} by the rotation eigenvalues $\pm 1, \pm i$, whereas symmetric WFs locating at $B,D$ are specified by the twofold rotation eigenvalues $\pm 1$. 
Henceforth, the ground state of an atomic insulator in a $\mathcal{C}_4$-symmetric crystal can be characterized by a set of ten integers~\cite{note1}
$N_{W;r}$, each of which denotes the number of occupied symmetric WFs located at site $W$ with rotation eigenvalue $r$. However, this set of integers does not yield a $\mathbb{Z}^{10}$ topological classification, since, taken separately, the  integers $N_{W;r}$ do not represent genuine topological invariants.

We illustrate this
by considering  a paradigmatic tight-binding model [see Fig.~\ref{fig:WannierC4}(a)] with four atomic sites in the unit cell at full filling. Considering first the longer-range hopping amplitudes $t_{2,3}$ to be much smaller than the nearest-neighbor hopping $t_1$, the system can be described in terms of four WFs centered at the Wyckoff position $A$, which yields 
$N_{A; \pm 1, \pm i} \equiv 1$ while $N_{B ; \pm 1} \equiv N_{C; \pm 1, \pm i} \equiv 0$. However, by continuously increasing the diagonal hopping $t_3$, the system can be more conveniently described in terms of four symmetric WFs centered at the Wyckoff position $C$, thus implying $N_{C; \pm 1, \pm i} \equiv 1$ and $N_{B ; \pm 1} \equiv N_{A; \pm 1, \pm i} \equiv 0$. Likewise, in the $t_2 \gg t_{1,3}$ regime the system can be described in terms of symmetric WFs centered at the twofold symmetric Wyckoff positions $B$ and $D$, in which case $N_{A; \pm 1, \pm i} \equiv N_{C; \pm 1, \pm i} \equiv 0$ while $N_{B ; \pm 1} \equiv N_{D ; \pm 1} \equiv 1$. Since these  continuous deformations in the hopping patterns do not change the topology of the insulating state, we find that the three configurations in Fig.~\ref{fig:WannierC4}(b) are all equivalent. 

To proceed further, we therefore define a subset of integers demanding their topological ``immunity"  against such continuous deformations. Since the latter simply correspond to $N_{W ; r} \rightarrow N_{W ; r} \pm 1$ $\forall~r$, we are led to define a new set of seven integers explicitly reading 
\begin{eqnarray}
\nu_{W ; \overline{r}}&=& -3 N_{W; \overline{r}} + \sum_{r \neq \overline{r}} N_{W ; r} \hspace{.4cm} \overline{r}=\pm 1, i~;W=A,C \nonumber \\ 
\nu_B&=&-N_{B;1} + N_{B; -1}. \label{eq:C4Wannierinvariants}
\end{eqnarray}
The reduced $\mathbb{Z}^7$ topological classification predicted by this integer set is in perfect agreement with the band structure combinatoric approach of Ref.~\cite{kru17}, where the topological invariants are related to the multiplicities $\Gamma_{r}$, $M_{r}$, and $X_{r}$  of the fourfold and twofold rotation eigenvalues $r$ at the high-symmetry (HS) points in the Brillouin zone \cite{note2}.  
This motivates us to find a one-to-one correspondence between the real-space invariants listed in Eq.~\eqref{eq:C4Wannierinvariants} and the rotation-symmetry labels of the band structure. It has a dual purpose: first, it proves that a change in the integers reported in Eq.~\eqref{eq:C4Wannierinvariants} is necessarily accompanied by a bulk gap closing-reopening; second, it allows to bypass the problem of constructing symmetric WFs. In particular, we obtain the following relations (see Ref.~\cite{notesup} for the other invariants):
\begin{align}
\nu_{A;1}&=\scalebox{0.88}[1]{$-3\Gamma_1-\Gamma_{-1}+2M_{-1}+X_{-1}+\dfrac{3}{2}[M_{-i}+M_i-\Gamma_i-\Gamma_{-i}]$}\nonumber \\
\nu_{C;1}&= 2[\Gamma_{-1}-M_{-1}]-X_{-1}+\dfrac{3}{2}[ \Gamma_i+\Gamma_{-i}]-\dfrac{M_i+M_{-i}}{2} \nonumber\\
\nu_B&=\dfrac{1}{2}[\Gamma_i+\Gamma_{-i}-M_i-M_{-i}].
\label{eq:C4Blochinvariants}
\end{align}

\begin{figure}[tbp]\centering
\includegraphics[width=.92\columnwidth]{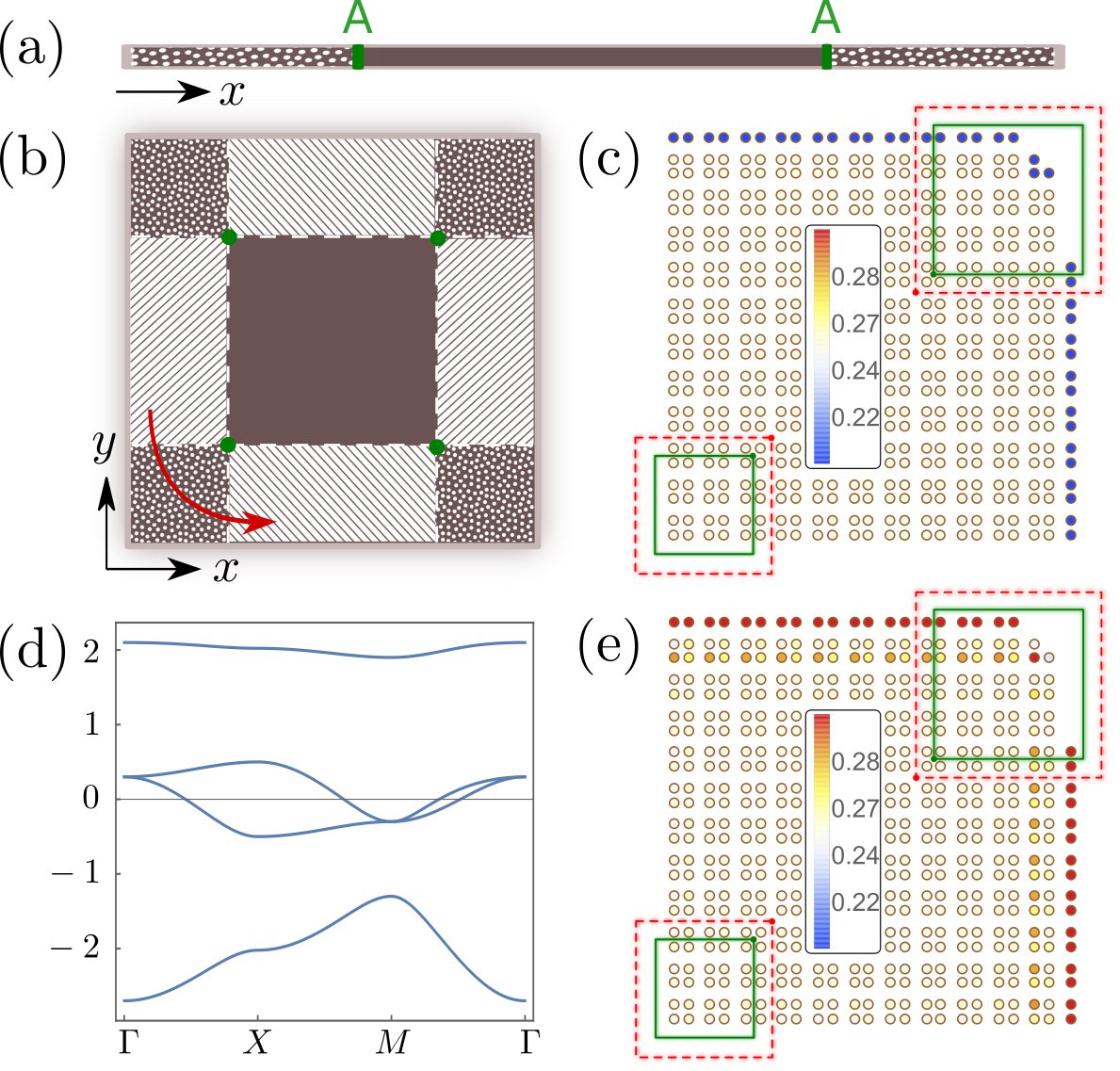}
\caption{(Color online) Sketch of a finite inversion-symmetric 1D chain (a), and a finite $\mathcal{C}_4$-symmetric 2D crystal (b). (c) Local charge distribution at quarter-filling for the model shown in Fig.~1(a) with $t_1=1$, $t_2=t_3=0$, and $E_F=-1.8$. (d) Bulk band structure for $t_1=1$, $t_2=0.2$, and $t_3=0.3$. Here $\Gamma=(0,0)$, $M=(\pi,\pi)$, and $X=(\pi,0)$. (e) Corresponding local charge distribution with  $E_F=-0.8$. The charge inside the red dashed (green solid) squares corresponds to $Q_A$ ($Q_C$). }
\label{fig:tbmodelc4}
\end{figure} 

\paragraph{Corner charges --} Having established the $\mathbb{Z}^7$ topological classification of  $\mathcal{C}_4$ symmetric crystals solely by virtue of symmetric WFs, we next show how this crystalline topology can be probed in systems with open geometries. We first recall that in a  inversion-symmetric insulating finite atomic chain the excess left (right) edge charge, defined as the fractional part of the total charge measured from $-\infty$ ($+\infty$) up to some reference point located sufficiently far away from the edges, is quantized to $0$ or $1/2$ if the reference point coincides with one of the two inversion centers of the chain [c.f. Fig.~\ref{fig:tbmodelc4}(a)]. This is because 
the excess left $Q_A^L$ and right $Q_{A}^R$ edge charges must be identical due to inversion symmetry. In addition, the total bulk charge contained between the two inversion reference centers  must be integer, which therefore implies $2 Q_{A}^L=2 Q_A^R = 0~\textrm{mod}~1$. Moreover (see Ref.~\cite{notesup}), the half-integer  value of the excess end charge can be immediately related to the $\mathbb{Z}_2$ part of the integer crystalline topological invariant for an inversion-symmetric insulating atomic chain.

Now we show that a very similar result holds for the corner charge of a $\mathcal{C}_4$-symmetric insulator. Let us consider the square geometry shown in Fig.~\ref{fig:tbmodelc4}(b) with four symmetry-related corners, and define the corner charge $Q_A$ as the fractional part of the charge contained in the region $\left(-\infty,j \right]\times\left(-\infty,j \right]$, with $j$ a sufficiently large  integer. This automatically constraints the reference point to coincide with a Wyckoff position $A$. $Q_A$ can be dubbed as a ``topological" bulk quantity if and only if its quantized value is insensitive to the microscopic details both at the corners and at the edges. The former condition is immediately verified if we concomitantly assume both the bulk and the edges to be completely insulating. To investigate the insensitivity to microscopic details at the edge, we consider the situation in which an edge potential is applied to, {\it e.g.}, the lower edge in Fig.~\ref{fig:tbmodelc4}(b). The induced charge flow from the corner to the edges, or {\it vice versa}, will result in a loss of quantization of the corner charge $Q_A$. However, if the applied edge potentials respect the fourfold rotational symmetry of the bulk crystal, the charge flow induced by the lower edge will be precisely compensated by the charge flow due to the left edge. Hence, the fourfold rotational symmetry of the crystal guarantees the bulk nature of the corner charge. The discretization of its value can be read off by considering that  $4 Q_A + 4 Q_{\textrm{edge}} = 0~\textrm{mod}~1$, where $Q_{\textrm{edge}}$ represents the edge charge contained in the dashed region of Fig.~\ref{fig:tbmodelc4}(b).  Since the latter is quantized to half-integer values \cite{mie17}, we find that the corner charge assumes values discretized in multiples of $1/4$ modulo an integer. More importantly,  we can immediately relate the corner charge to the formerly defined $\mathbb{Z}$ integer invariants using that $Q_A = \sum_{r} N_{A; r}/4 \equiv \nu_{A; \bar{r}} / 4$ modulo an integer $\forall~\bar{r}$. These relations can be rationalized by considering that for a fourfold rotational-symmetric insulator in the atomic limit, {\it i.e.} where all hopping amplitudes connecting atomic sites are set to zero, the fractional part of the corner charge comes only about WFs centered at the reference point $\left\{j,j\right\}$, each of which contributes with a quarter of the electronic charge.  By invoking the principle of adiabatic continuity, this result holds also when switching on the hopping amplitudes provided the band gap does not close and reopen. A similar analysis can be performed for the corner charge with a reference point $\left\{j+1/2, j+1/2\right\}$ corresponding to a Wyckoff position $C$. Moreover, for corners where the reference point is chosen at the twofold rotational symmetric Wyckoff positions $B$ and $D$, it is possible to show that the sum of the charge contained in two corner partners is quantized in multiples of half of the electronic charge~\cite{notesup}. 
All in all, we therefore find that corner charges are able to probe a $\mathbb{Z}_4 \otimes \mathbb{Z}_4 \otimes \mathbb{Z}_2$ part of the general $\mathbb{Z}^7$ crystalline topology. 

To corroborate our findings, we next present explicit calculations for the tight-binding model shown in Fig.~\ref{fig:WannierC4}(a). Fig.~\ref{fig:tbmodelc4} (c) shows the charge distribution at one-quarter filling setting the hopping amplitudes $t_2~\equiv~t_3~\equiv0$. In this case the system consists of uncoupled molecules centered at Wyckoff position $A$. Consequently the bulk band structure is made out of flat bands with energies $\pm 2, 0$. For the regular lower left corner in Fig.~\ref{fig:tbmodelc4}(c) visual inspection immediately reveals that the corner charge $Q_A=1/4$ while $Q_C=0$, which is in perfect agreement with the fact that $\nu_{A; \bar{r}}=1~\textrm{mod}~4$ and $\nu_{C; \bar{r}}=0~\textrm{mod}~4$. Notice that these values of the corner charges are insensitive to the specific termination as long as the edges are related by the fourfold symmetry [c.f. the upper right corner in Fig.~\ref{fig:tbmodelc4}(c)]. More importantly, the quantization of the corner charge is preserved even for $t_2 \neq t_3 \neq 0$, in which case the bulk band acquire a full dispersion [c.f. Fig.~\ref{fig:tbmodelc4}(d)] and the electronic charge density in the corner region is manifestly inhomogeneous [c.f. Fig.~\ref{fig:tbmodelc4}(e)].

\begin{figure}[tbp]\centering
\includegraphics[width=.95\columnwidth]{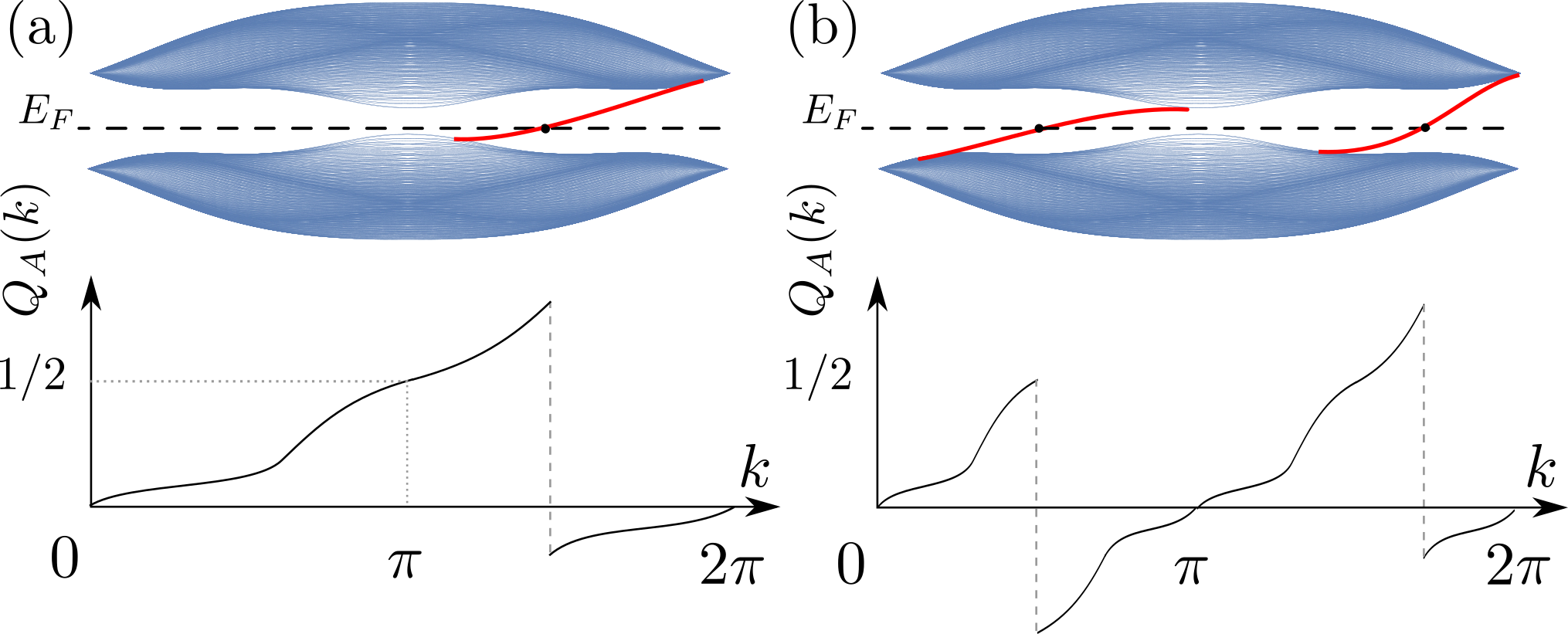}
\caption{(Color online) End-charge flow patterns for inversion-symmetric (a) odd- and (b) even-integer Chern insulators. In both cases, the edge charge is an odd function in the momentum $k$ up to integer jumps.}
\label{fig:chargeflow}
\end{figure}

\paragraph{Chiral hinge states in rotoinversion symmetric crystals --} 
Next, we show how the topological invariants underpinning the presence of quantized corner charges can be used to define the bulk $\mathbb{Z}_2$ topological invariant of a second-order topological insulator. Let us first consider the aforementioned example of a 
two-dimensional Chern insulating state in an inversion-symmetric crystal put in a ribbon geometry. Assuming the translational symmetry to be preserved in the $\hat{x}$ direction, one can view the full system as a collection of one-dimensional finite chains with the momentum $k_x$ playing the role of an external parameter. Inversion symmetry of the two-dimensional crystal implies that the right edge charge at $k_x$ is identical to the left edge charge at $-k_x$. This, in turns, implies the constraint for the single edge charge $Q_A(k_x)=-Q_A(-k_x)~\textrm{mod}~1$, i.e. $Q_A(k_x)$ is an odd function up to integer jumps, which correspond to edge states crossing the Fermi level $E_F$. Moreover, this guarantees the quantization of the edge charge along the inversion-symmetric lines $k_x=0,\pi$. With these relations, we can identify two ``topologically" distinct states of the original two-dimensional crystal. In fact, for $Q_{A}(0)-Q_{A}(\pi)=1/2~\textrm{mod}~1$ an odd number of in-gap states must cross the Fermi level at each edge [c.f. Fig.~\ref{fig:chargeflow}(a)] during an adiabatic cycle of $k_x \in \left[0, 2 \pi \right]$. On the contrary, an even number of in-gap states will cross the Fermi level for $Q_{A}(0)-Q_{A}(\pi)=0~\textrm{mod}~1$ [c.f. Fig.~\ref{fig:chargeflow}(b)]. As a result, we find that the crystalline topology of the two inversion-symmetric chains at $k_x=0,\pi$ is able to diagnose  the  Chern number parity of the 2D crystal.

Let us now exploit a similar connection for the quantized corner charges and consider a three-dimensional crystal with a fourfold rotoinversion symmetry [see Ref.~\cite{notesup} for other point-group symmetries] ${\cal S}_4 = {\cal C}_4 \times {\cal M}_z$, where ${\cal M}_z$ indicates the mirror symmetry in the $\hat{z}$ direction. As before, we consider the bulk three-dimensional Hamiltonian as a collection of two-dimensional Hamiltonians ${\cal H}({k_z})$ parametrized by the momenum $k_z$. The fourfold rotoinversion symmetry immediately implies that the two-dimensional Hamiltonians ${\cal H}(0), {\cal H}(\pi)$ both inherit the fourfold rotoinversion symmetry ${\cal S}_4$. Therefore, the three-dimensional crystal can be classified topologically by $\mathbb{Z}^{14}$ \cite{notecnmz}. These fourteen invariants, however, are not all independent when requiring a full bulk bandgap. This is due to the fact that the collection of two-dimensional Hamiltonians all possess a twofold rotational symmetry ${\cal C}_2 = \left({\cal S}_4\right)^2$. 
In order to prevent bandgap closing along the HS lines $(0,0,k_z)$, $(\pi,0,k_z)$, $(0,\pi,k_z)$, and $(\pi,\pi,k_z)$, the multiplicities of the twofold rotational symmetry eigenvalues must therefore remain constant. These four constraints imply that the two-dimensional topological crystalline integer invariants $\nu_{A (C); \bar{r}}(k_z = 0)= \nu_{A (C); \bar{r}}(k_z=\pi)~\textrm{mod}~2 $ and $\nu_{B}(k_z=0)\equiv \nu_{B}(k_z=\pi)$. 

\begin{figure}[tbp]
\centering
\includegraphics[width=.95\columnwidth]{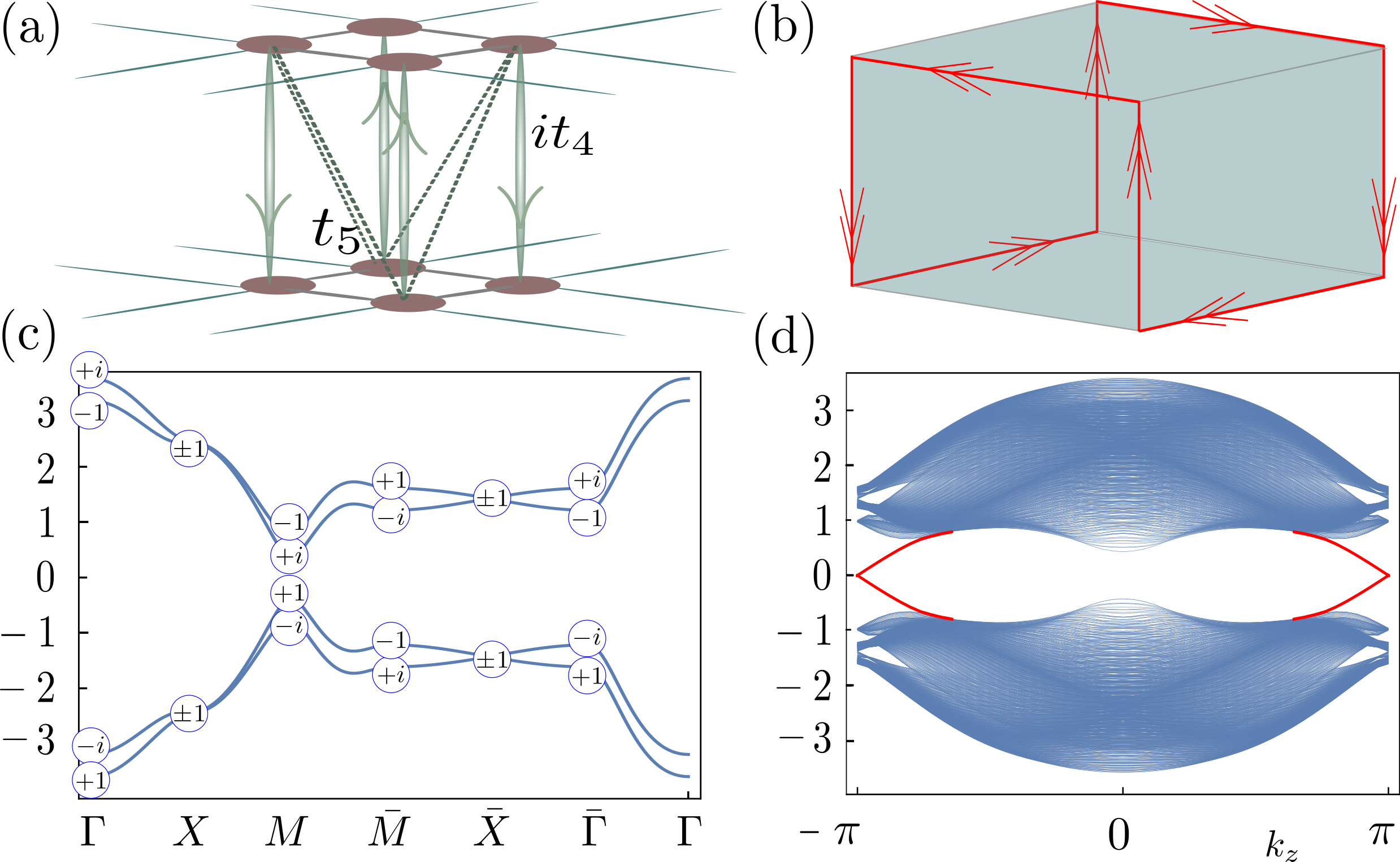}
\caption{(Color online) (a) Close-up of the 3D crystal. (b) Chiral hinge state pattern in a non-trivial $\mathcal{S}_4$-symmetric insulator. (c) Bulk band structure along HS lines, with the corresponding symmetry labels. HS points with (without) bar lie in the $k_z=\pi$ ($k_z=0$) plane. Here, we have used $t_1=t_5=0.7$, $t_2=1$, $t_3=.2$, and $t_4=0.5$. (d) Band structure with periodic boundary conditions in the $\hat{z}$-directions and open in the $\hat{x}$ and $\hat{y}$ directions.}
\label{fig:chiralhingestates}
\end{figure}

With these constraints in our hands, we can now analyze the flow of the corner charge $Q_{A}$  as the momentum $k_z$ completes an adiabatic cycle. The fourfold rotoinversion symmetry ${\cal S}_4$ ensures $Q_A(k_z)=-Q_A(-k_z)\textrm{ mod }1/2$, which is in agreement with the quantization of the corner charge in multiples of one quarter of the electronic charge in the planes $k_z=0$ and $k_z=\pi$ \cite{notesup}. As a result, we can only distinguish between two ``topologically" distinct states. In fact, for 
$Q_A(0)=Q_A(\pi)~\textrm{mod}~1$ the flow of the corner charge as $k_z$ makes an adiabatic cycle will be qualitatively similar to Fig.~\ref{fig:chargeflow}(b), and the hinges of the three-dimensional crystal will host an even number of chiral states. On the contrary, for $Q_A(0)=Q_A(\pi)\pm 1/2~\textrm{mod}~1$, and similarly to Fig.~\ref{fig:chargeflow}(a), an odd number of chiral hinge states will cross the Fermi level, thereby leading to a pattern of chiral hinge modes schematically shown in Fig.~\ref{fig:chiralhingestates}(b). Therefore, the difference in the integer invariants $\nu_{A; \bar{r}}$ at $k_z=0,\pi$ provides the bulk $\mathbb{Z}_2$ topological invariants for a fourfold rotoinversion symmetry-protected second-order topological insulator. 

Two remarks are in order here. First, the difference between the integer invariants $\nu_{C; \bar{r}}$ at $k_z=0,\pi$ equals the difference in the $A$ invariants, thereby implying the existence of a single $\mathbb{Z}_2$ bulk topological invariant \cite{noteNANC}. Second, contrary to the example of an inversion-symmetric Chern insulators where the crystalline symmetry only provides us with a criterion for the Chern number parity \cite{tur10,fan12}, here the fourfold rotoinversion ${\cal S}_4$ represents the fundamental symmetry protecting the presence of chiral hinge states. It therefore plays the same stabilizing role  time-reversal symmetry plays in two- and three-dimensional topological insulators in class AII. 

We finally corroborate our findings by considering a tight-binding model corresponding to a three-dimensional stack of the two-dimensional crystal shown Fig.~\ref{fig:WannierC4}(a). The layers are connected by purely imaginary hoppings $\pm i t_4$, as well as real  diagonal next-nearest neighbor hoppings $t_5$ [c.f. Fig.~\ref{fig:chiralhingestates}(a)]. Note that these interlayer hoppings break separately the $\mathcal{C}_4$ and $\mathcal{M}_z$ symmetries, but respect their combination. 
Finally, we have included a $\pi$-flux threading each plaquette surrounding Wyckoff positions $A$, $B$, $C$ and $D$.  
At half-filling, a direct computation of the two-dimensional crystalline invariants using Eq.~\ref{eq:C4Blochinvariants} on the $k_z=0,\pi$ planes [c.f. Fig.~\ref{fig:chiralhingestates}(c)] yields $\nu_{A; \bar{r}}(k_z=0)=2~\textrm{mod}~4$ and  $\nu_{A; \bar{r}}(k_z=\pi)=0~\textrm{mod}~4$, see Fig.~\ref{fig:chiralhingestates}(c). Our $\mathbb{Z}_2$ topological criterion then predicts the presence of chiral hinge states when considering the system in an open geometry.  
This is precisely what we find by diagonalizing the corresponding Hamiltonian with open boundary conditions in the $\hat{x}$ and $\hat{y}$ direction: two chiral hinge states going upwards and two chiral hinge states going downwards transverse the bulk band gap of the system, see Fig.~\ref{fig:chiralhingestates}(d), in perfect agreement with the hinge mode pattern shown in Fig.~\ref{fig:chiralhingestates}(b). 

\paragraph{Conclusions --} To sum up, we have characterized the bulk topology of second-order topological insulators with anomalous chiral hinge modes protected by (roto)inversion symmetries. The corresponding bulk $\mathbb{Z}_2$ topological invariant, defined in terms of the occupied symmetric hybrid Wannier functions, 
can be equally expressed using the crystalline symmetry labels in the bulk band structure representation. 
Therefore, our invariant can be straightforwardly computed not only using tight-binding models but also in density functional theory calculations where higher-order topological candidate materials can be identified. 
A promising future direction is to extend the methodology proposed in this work to time-reversal symmetric systems. 

\begin{acknowledgments}
C.O. acknowledges support from a VIDI grant (Project 680-47-543) financed by the Netherlands Organization for Scientific Research (NWO). This work is part of the research programme of the Foundation for Fundamental Research on Matter (FOM), which is part of the Netherlands Organization for Scientific Research (NWO).
\end{acknowledgments}

\appendix
\section{Supplemental Material for Higher-order topological insulators protected by (roto)inversion symmetries}
\section{Topology of inversion-symmetric 1D crystals}
In this section we discuss the topology of 1D inversion-symmetric crystals. We show that symmetry-adapted Wannier functions give rise to two topological $\mathbb{Z}$-invariants, and we will link these to the momentum space invariants.

Let us denote the Wannier functions for a one-dimensional inversion-symmetric chain in the compact form $w_{m; \alpha}(x)=w_\alpha(x-m)$, where $\alpha$ is the Wannier index while $m$ is the unit-cell index (the lattice constant has been set to $a=1$ for simplicity). 
The inversion symmetry of the chain implies the existence of two inversion centers within the unit cell, which, with a proper choice of the origin, can be fixed to $x=0$ (Wyckoff position $A$) and $x=1/2$ (Wyckoff position $B$). 

For a symmetric Wannier function centered at the Wyckoff position $A$, the following transformation properties apply: 
$$w_{m; \mu, A}(-x) \equiv \pm w_{-m ; \mu, A}(x),$$ 
where $\pm$ refers to Wannier functions that are respectively even and odd under inversion, and we have explicitly split the Wannier index $\alpha$ into a site index and  an orbital index $\mu$. Symmetric Wannier functions centered at $B$ instead transform as 
$$w_{m; \mu, B}(-x) \equiv \pm w_{-m-1 ; \mu, B}(x).$$ 
These different transformation properties allow us to classify crystalline inversion-symmetric 
one-dimensional insulators with the set of four integers $N_{A(B);\pm 1}$, {\it i.e.} the total number of occupied symmetric Wannier functions centered at $A$($B$) that are even and odd under inversion.  
As explained in the main text for $\mathcal{C}_4$ rotational-symmetric crystals, $N_{A(B);\pm 1}$ do not represent genuine topological invariants.

\begin{figure}[t]\centering
\includegraphics[width=.9\textwidth]{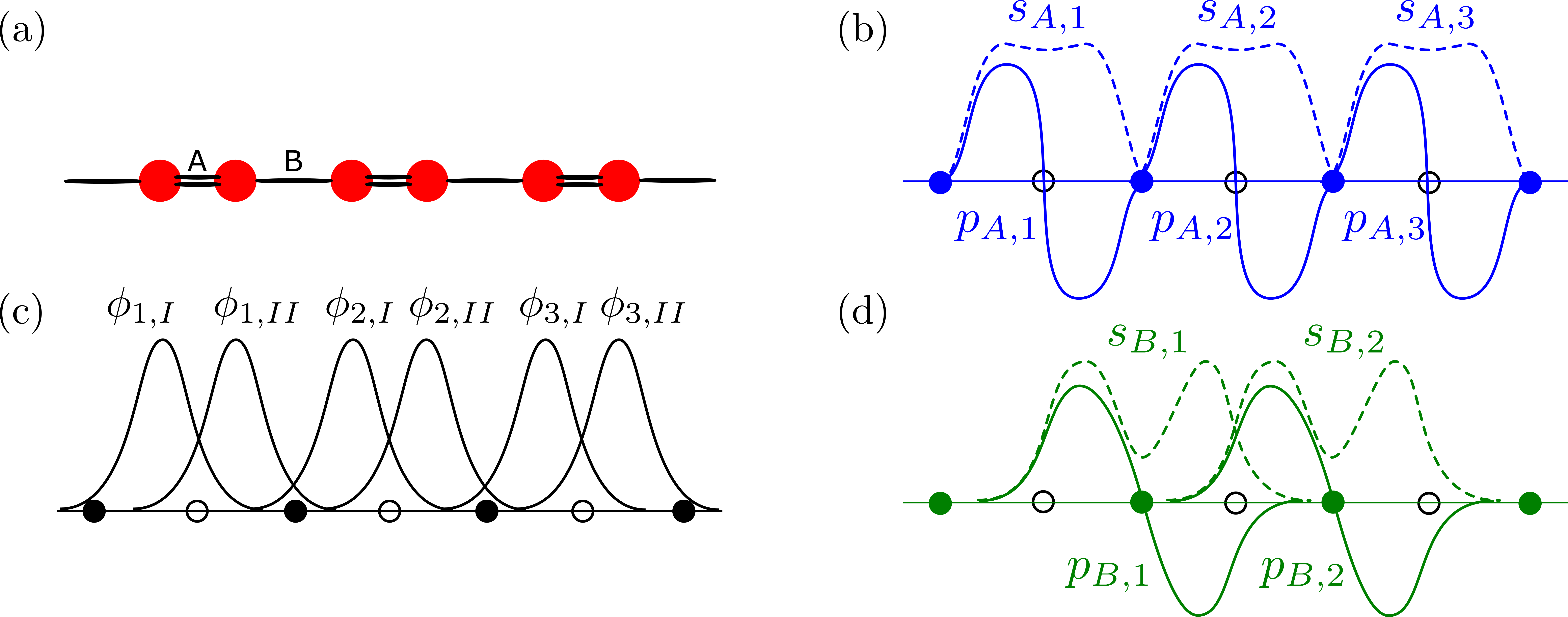}
\caption{(a) Sketch of an Su-Schrieffer-Heeger chain. (b) Symmetry-adapted Wannier functions with respect to A. (c). Wannier functions centered at the atomic sites. (d) Symmetry-adapted Wannier functions with respect to B.}
\label{fig:ssh}
\end{figure}

Let us consider the Su-Schrieffer-Heeger model for spinless electrons [see Fig.~\ref{fig:ssh}(a),(b)] at integer filling. The ground state of this model can be expressed as a Slater determinant involving s-like and p-like orbitals centered at the inversion center $A$ [c.f. Fig.~\ref{fig:ssh}(c)], whose Wannier functions we denote by $w_{m; s,A}(x)$ and $w_{m; p, A}(x)$ respectively. This implies that the associated set of integers defined above simply reads $N_{A;1}=N_{A;-1}=1$, and $N_{B;1}=N_{B;-1}=0$. However, this set of integers is not unique. In fact, starting out from the symmetric Wannier functions $w_{m; s/p ,A}(x)$, it is possible to define new symmetric Wannier functions centered at $B$ [c.f. Fig.~\ref{fig:ssh}](d) using the following relations 
\begin{eqnarray*}
w_{m; s, B}(x)&=&\frac{1}{2}\left[w_{m;s, A}(x)+w_{m+1; s, A}(x)+ \right. \\
& & \left.+ w_{m;p,A}(x)-w_{m+1;p,A}(x)\right]\\
w_{m;p,B}(x)&=&\frac{1}{2}\left[-w_{m;s,A}(x)+w_{m+1;s,A}(x) \right. \\
& & \left. -w_{m;p,A}(x)-w_{m+1;p,A}(x)\right]
\end{eqnarray*}
For these new Wannier functions, the set of integers clearly reads $N_{A;1}=N_{A;-1}=0$ and $N_{B;1}=N_{B;-1}=1$.

The genuine topological invariants 
correspond to the difference between even and odd Wannier functions centered at the two Wyckoff positions, and thus read: 
\begin{align}
\nu_A=-N_{A;1}+N_{A;-1} \hspace{.5cm}; \hspace{.5cm} \nu_B=-N_{B;1}+N_{B;-1} .
\end{align}
To prove 
that these integers are indeed topological invariants, we will now relate them to the symmetry labels in the band representation. We recall that the symmetry indicators of the band structure of a one-dimensional inversion-symmetric crystal correspond to the multiplicity of the even and odd eigenvalues of the inversion operator (with respect to inversion center $A$) at the BZ center $\Gamma=0$, which we dub $\Gamma_{1}$ and $\Gamma_{-1}$, and at the BZ edge $X=\pi$ , dubbed $X_{1}$ and $X_{-1}$. These symmetry labels are not independent since the existence of a band gap together with the continuity of the energy bands yields global constraints, known as compatibility relations, explicitly reading $\Gamma_{1} + \Gamma_{-1} \equiv X_{1} + X_{-1} \equiv N_F$, with $N_F$ indicating the total number of occupied bands. 
To proceed further, we relate the two sets of integers stipulating the ansatz $\nu_{A(B)}= \gamma_{A(B)}^{-1}\Gamma_{-1}+\gamma_{A(B)}^1\Gamma_1+x_{A(B)}^{-1}X_{-1}$, where we used that $X_{1}$ can be expressed in terms of the remaining symmetry labels via the compatibility relations. The linearity in the ansatz above is required by the fact that the topological invariant of two uncoupled systems must be additive. Next, we will determine the six unknowns $\gamma_{A(B)}^{-1}, \gamma_{A(B)}^1, x_{A(B)}^{-1}$ using the following constraints: 
\begin{enumerate}
\item  $\nu_{A(B)}=-1$, if we have a single even Wannier function centered a $A(B)$,
\item  $\nu_{A(B)}=+1$, if we have a single odd Wannier function centered a $A(B)$,
\item $\nu_{A(B)}=0$, if we have a single Wannier function centered at $B(A)$.
\end{enumerate}
To do so, we analyze the inversion eigenvalues 
of the Bloch states  $\psi_{q; \mu, \sigma}(x)= \sum_m e^{i q m} w_{m; \mu, \sigma}(x)$. Using the transformation properties of the symmetric Wannier functions, one finds
\begin{equation*}
\left\{ \begin{array}{l} \psi_{q; \mu, A}(-x) = \pm \, \psi_{-q; \mu, A}(x) \\ \\
 \psi_{q; \mu, B}(-x) = \pm \mathrm{e}^{i q} \,  \psi_{-q; \mu, A}(x),
\end{array} \right.
\end{equation*}
where the plus (minus) sign refers to symmetric Wannier functions that are even (odd) under inversion. 
In Table~\ref{tbl:1dinv} we list the inversion eigenvalues of the Bloch states corresponding to even and odd Wannier functions centered at the two Wyckoff positions.
\begin{table}[t!]
\begin{tabular}{|c|c||c|c|c|c|}
\hline
~~~~$W$~~~~&~~~~$\lambda$~~~~&~~~~$\Gamma_1$~~~~&~~~~$\Gamma_{-1}$~~~~&~~~~$X_{-1}$~~~~\\
\hline\hline
A& $1$&1&0&0\\
& $-1$&0&1&1\\
\hline
B& $1$&1&0&1\\
 &$-1$&0&1&0\\
\hline
\end{tabular}
\caption{Multiplicities at inversion-invariant momenta for symmetry adapted Wannier functions at Wyckoff position $W$ with inversion eigenvalue $\lambda$. }\label{tbl:1dinv}
\end{table}
With this, the constraints listed above yield the following relations 
\begin{equation*}
\left\{
\begin{array}{l} 
-1=\gamma_A^{1}= \gamma_B^1+x^{-1}_B \\
1=\gamma_A^{-1}+x^{-1}_A=\gamma^{-1}_B \\
0=\gamma_A^1+x_A^{-1}=\gamma_A^{-1}=\gamma^{-1}_B+x_B^{-1} = \gamma^1_B ,
\end{array}
\right.
\end{equation*}
which thereafter yield the expression for the topological invariants in terms of the inversion eigenvalues 
\begin{eqnarray}
\nu_A =& \Gamma_{-1} -X_1 & = -\Gamma_{1} + X_{-1} \\
\nu_B =& -\Gamma_1 + X_1 & = \Gamma_{-1} - X_{-1} 
\end{eqnarray} 
All in all, the equations above serve a dual purpose: first they prove that $\nu_{A,B}$ are genuine topological invariants, since the inversion-symmetry eigenvalues can only change upon closing and reopening the band gap. Second, we have at hands an efficient strategy to bypass the problem of finding symmetry-adapted Wannier functions. 

\section{Edge charge of inversion-symmetric 1D crystals}
The 1D invariants $\nu_{A}$ and $\nu_B$ introduced above have a well-defined meaning. Let us consider a finite inversion-symmetric insulator in the atomic limit, with $A$ as its center (the same considerations apply to $B$). Then, we find that the total number of states that is odd under inversion minus the total number of states that is even with respect to the center of the chain is precisely equal to $\nu_A$. By invoking the principle of adiabatic continuity, we find that this statement holds in general.

Moreover, the $\mathbb{Z}_2$-part of $\nu_A$ pinpoints the value of the left and right edge charge, measured with respect to $A$: if $\nu_A$ is odd (even), there will be an odd (even) number of electrons on the chain, which implies that the left and right excess edge charge are semi-integer (integer), when measured with respect to $A$  [see the main part of the manuscript]. Although this excess charge only probes the $\mathbb{Z}_2$-part, it is very robust in nature. Namely, even if the finite chain itself is not inversion-symmetric, as is typically the case, still the excess charges will be quantized for the very simple reason that charge cannot flow away from the edge as long as the bulk is insulating.  

\section{Topology of 2D rotation-symmetric crystals}
In this section we discuss the topology of 2D rotation-symmetric crystals with zero Chern number. We provide a dictionary between the real-space topology expressed in terms of symmetry-adapted Wannier functions and the symmetry-labels in the band representation. We discuss in detail the topology of $\mathcal{C}_4$-symmetric crystals, after which we simply state the results for $\mathcal{C}_2$, $\mathcal{C}_3$, and $\mathcal{C}_6$-symmetric crystals.
\begin{figure}[h!]\centering
\includegraphics[width=.8\textwidth]{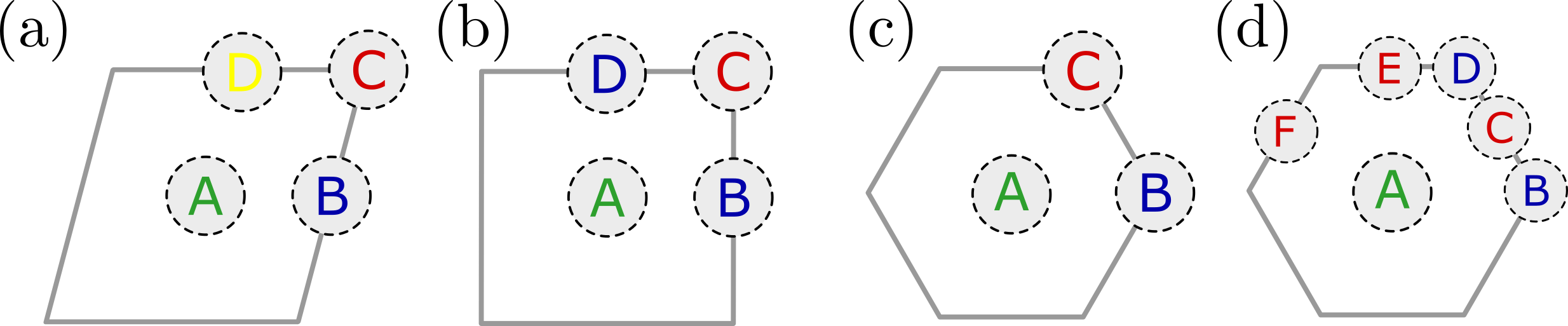}
\caption{Wyckoff positions in rotation-symmetric crystals. (a) $\mathcal{C}_2$-symmetric crystal. (b) $\mathcal{C}_3$-symmetric crystal. (c) $\mathcal{C}_4$-symmetric crystal. (d) $\mathcal{C}_6$-symmetric crystal. Wyckoff positions related to each other by the rotation symmetry have the same color.}
\label{fig:wyckoff}
\end{figure} 
\subsection{Integer crystalline topological invariants in $\mathcal{C}_4$-symmetric crystals}
In a $\mathcal{C}_4$-symmetric crystal, there are two Wyckoff positions, $A$ and $C$ that are invariant under four-fold rotations, and two Wyckoff positions, $B$ and $D$ that are separately invariant under two-fold rotations, and transform into eachother under a fourfold rotation, see Fig.~\ref{fig:wyckoff}(b). The Wannier functions are symmetry-adapted if they satisfy the following transformation properties:
\begin{align}
w_{\mathbf{m};\mu,A}(R\mathbf{x})&=r_{\mu,A}w_{R^{-1}\mathbf{m};\mu,A}(\mathbf{x}),\label{eq:symWannier1}\\
w_{\mathbf{m};\mu,C}(R\mathbf{x})&=r_{\mu,C}w_{R^{-1}\mathbf{m}-\mathbf{e}_2;\mu,C}(\mathbf{x}),\label{eq:symWannier2}\\
w_{\mathbf{m};\mu,B}(R\mathbf{x})&=r_{\mu,B/D}w_{R^{-1}\mathbf{m}-\mathbf{e}_2;\mu,D}(\mathbf{x}),\label{eq:symWannier3}\\
w_{\mathbf{m};\mu,D}(R\mathbf{x})&=w_{R^{-1}\mathbf{m};\mu,B}(\mathbf{x}).\label{eq:symWannier4}
\end{align}
where $R$ denotes the rotation matrix, given by
\begin{align*}
R&=\begin{pmatrix}
0&-1\\1&0
\end{pmatrix},
\end{align*}
$\mathbf{e}_2=(0,1)$, and the rotation eigenvalues satisfy $r_{\mu,A}^4=r_{\mu,C}^4=r_{\mu,B/D}^2=1$. Note that this corresponds to a fourfold rotation around $A$. Given these Wannier functions we can define a set of integer $N_{A;1},\ldots,N_{A;-i},N_{C;1},\ldots,N_{C;-i},N_{B;1}$, $N_{B;-1}, N_{D;1}$, and $N_{D;-1}$ where $N_{A,\lambda}$ ($N_{C,\lambda}$) denotes the number of occupied fourfold symmetric Wannier functions centered at $A$ ($C$) with rotation eigenvalue $\lambda$, and $N_{B(D),\lambda}$ denotes the number of occupied twofold symmetric Wannier functions centered at $B$ ($D$) with eigenvalue $\lambda$. Note that the fourfold rotational symmetry implies $N_{B;\pm1}=N_{D;\pm1}$. These integers, in turn, can be used to define the invariants $\nu_{A; \bar{r}}$, $\nu_{C; \bar{r}}$, and $\nu_{B}\equiv \nu_D$ defined in the main part of the manuscript. 
\begin{table}[t]
\begin{tabular}{|c|c||c|c|c|c|c|c|c|c|}
\hline
~~$W$~~&~~$\lambda$~~&~$\Gamma_1$~&~$\Gamma_{i}$~&~$\Gamma_{-1}$~&~$\Gamma_{-i}$~&~$M_{i}$~&~$M_{-1}$~&~$M_{-i}$~&~$X_{-1}$~\\
\hline\hline
A&1&1&0&0&0&0&0&0&0\\
&i&0&1&0&0&1&0&0&1\\
&-1&0&0&1&0&0&1&0&0\\
&-i&0&0&0&1&0&0&1&1\\
\hline
C&1&1&0&0&0&0&1&0&1\\
&i&0&1&0&0&0&0&1&0\\
&-1&0&0&1&0&0&0&0&1\\
&-i&0&0&0&1&1&0&0&0\\
\hline
B+D&1&1&0&1&0&1&0&1&1\\
&-1&0&1&0&1&0&1&0&1\\
\hline
\end{tabular}
\caption{Multiplicities of the rotation eigenvalues at $\mathcal{C}_4$- and $\mathcal{C}_2$-invariant momenta for symmetry-adapted Wannier orbitals at Wyckoff position $W$ and with rotation eigenvalue $\lambda$.}\label{tbl:c4multi}
\end{table}

We next make the following ansatz:
\begin{align*}
\nu_{W,\lambda}&=\gamma_{1}\Gamma_1+\gamma_{i}\Gamma_i+\gamma_{-1}\Gamma_{-1}+\gamma_{-i}\Gamma_{-i}\\
&+m_{i}M_i+m_{-1}M_{-1}+m_{-i}M_{-i}+x_{-1}X_{-1}
\end{align*}
with $W=A,C,B$. Moreover, $\Gamma_\lambda$ ($M_\lambda$) denotes the number of occupied bands at $\Gamma=(0,0)$ ($M=(\pi,\pi)$) with fourfold rotation eigenvalue $\lambda$, $X_{-1}$ is the number of states at $X=(\pi,0)$ that is odd under a twofold rotation, and $\gamma_{\mu}$, $m_{\mu}$, and $x_{\mu}$ are coefficients that we need to solve for. Let us shortly motivate this ansatz. Similarly to a one-dimensional atomic chain, 
we consider a function that is linear for the very simple reason that the numbers we have defined are additive. Second, we have not included the multiplicities $X_1$ and $M_1$, since we have used the following identity $\Gamma_1+\Gamma_i+\Gamma_{-1}+\Gamma_{-i}=M_1+M_i+M_{-1}+M_{-i}=X_1+X_{-1}$, which corresponds to the compatibility relation for at two-fold rotational symmetric crystal. Third, we have not included the multiplicities at $Y$ since these are identical to the multiplicities at $X$. In Table~\ref{tbl:c4multi} we list the multiplicities of the rotation eigenvalues for a single symmetry-adapted Wannier function centered at one of the four Wyckoff positions. These values have been obtained by Fourier transforming Eqs.~\eqref{eq:symWannier1}-\eqref{eq:symWannier4}. Using these values, we can solve for the eight unknowns of each integer topological invariant. We find the following relations: 
\begin{align*}
\nu_{A;1}&= -3\Gamma_1 -\dfrac{3}{2}\Gamma_i -\Gamma_{-1} -\dfrac{3}{2}\Gamma_{-i}+\dfrac{3}{2}M_i+2M_{-1}+\dfrac{3}{2}M_{-i}+X_{-1},\\
\nu_{A;i}&=\Gamma_1-\dfrac{1}{2}\Gamma_i+\Gamma_{-1}+\dfrac{3}{2}\Gamma_{-i}-\dfrac{3}{2}M_i+\dfrac{1}{2}M_{-1}-X_{-1},\\
\nu_{A;-1}&=\Gamma_1+\dfrac{1}{2}\Gamma_i-\Gamma_{-1}+\dfrac{1}{2}\Gamma_{-i}-\dfrac{1}{2}M_i-2M_{-1}-\dfrac{1}{2}M_{-1}+X_{-1},\\
\nu_{C;1}&=\dfrac{3}{2}\Gamma_i+2\Gamma_{-1}+\dfrac{3}{2}\Gamma_{-i}-\dfrac{1}{2}M_i-2M_{-1}-\dfrac{1}{2}M_{-i}-X_{-1},\\
\nu_{C;i}&=-\dfrac{3}{2}\Gamma_i +\dfrac{1}{2}\Gamma_{-i} +\dfrac{1}{2}M_{i}-\dfrac{3}{2}M_{-i}+X_{-1},\\
\nu_{C;-1}&=-\dfrac{1}{2}\Gamma_i-2\Gamma_{-1}-\dfrac{1}{2}\Gamma_{-i}+\dfrac{3}{2}M_{i}+2M_{-1}+\dfrac{3}{2}M_{-i}-X_{-1},\\
\nu_B&=\dfrac{1}{2}\Gamma_i+\dfrac{1}{2}\Gamma_{-i}-\dfrac{1}{2}M_i-\dfrac{1}{2}M_{-i}.
\end{align*}

\subsection{Integer crystalline topological invariants in $\mathcal{C}_2$-symmetric crystals}
In a 2D $\mathcal{C}_2$-symmetric crystal we find four $\mathcal{C}_2$-symmetric Wyckoff positions per unit cell. These are labeled as $A, B, C, $ and $D$, see Fig.~\ref{fig:wyckoff}(a). The Wannier functions are symmetry-adapted if they satisfy the following properties:
\begin{align}
w_{\mathbf{m};\mu,A}(-\mathbf{x})&=r_{\mu,A}w_{-\mathbf{m};\mu,A}(\mathbf{x}),\label{eq:symWannier1c2}\\
w_{\mathbf{m};\mu,B}(-\mathbf{x})&=r_{\mu,B}w_{-\mathbf{m}-\mathbf{e}_1;\mu,B}(\mathbf{x}),\label{eq:symWannier2c2}\\
w_{\mathbf{m};\mu,C}(-\mathbf{x})&=r_{\mu,C}w_{-\mathbf{m}-\mathbf{e}_1-\mathbf{e}_2;\mu,C}(\mathbf{x}),\label{eq:symWannier3c2}\\
w_{\mathbf{m};\mu,D}(-\mathbf{x})&=r_{\mu,D}w_{-\mathbf{m}-\mathbf{e}_2;\mu,D}(\mathbf{x}),\label{eq:symWannier4c2}
\end{align}
where $\mathbf{e}_1=(1,0)$ and the rotation eigenvalues satisfy $r^2_{\mu,A}=r^2_{\mu,B}=r^2_{\mu,C}=r^2_{\mu,D}=1$. Given these Wannier functions, let $N_{W;\pm1}$ denote the number of occupied two-fold rotation symmetric Wannier functions centered at $W$ with rotation eigenvalue $\pm 1$. Out of these eight integers we can extract four $\mathbb{Z}$ invariants given by
\begin{align*}
\nu_A&=-N_{A;1}+N_{A;-1},\\
\nu_B&=-N_{B;1}+N_{B;-1},\\
\nu_C&=-N_{C;1}+N_{C;-1},\\
\nu_D&=-N_{D;1}+N_{D;-1}.\\
\end{align*}
We now follow the strategy laid out in the main text and above to show that these are indeed proper invariants. Therefore, we relate them to the band structure multiplicities $\Gamma_{\pm 1}$, $X_{- 1}$, $Y_{- 1}$, and $M_{- 1}$. We simply state the results:
\begin{align*}
\nu_A&=-\Gamma_1-\frac{1}{2}\Gamma_{-1}+\frac{1}{2}X_{-1}+\frac{1}{2}Y_{-1}+\frac{1}{2} M_{-1},\\
\nu_B&=\frac{1}{2}\Gamma_{-1}-\frac{1}{2}X_{-1}+\frac{1}{2} Y_{-1}-\frac{1}{2} M_{-1},\\
\nu_C&=\frac{1}{2}\Gamma_{-1}-\frac{1}{2}X_{-1}-\frac{1}{2} Y_{-1}+\frac{1}{2} M_{-1},\\
\nu_D&=\frac{1}{2}\Gamma_{-1}+\frac{1}{2}X_{-1}-\frac{1}{2} Y_{-1}-\frac{1}{2} M_{-1}.
\end{align*}
These results have been obtained by using the multiplicities of the rotation eigenvalues for single symmetry-adapted Wannier functions centered at one of the Wyckoff positions, see Table~\ref{tbl:c2multi}.
\begin{table}[h!]
\begin{tabular}{|c|c||c|c|c|c|c|c|c|c|}
\hline
~~$W$~~&~~~$\lambda$~~~&~~~$\Gamma_1$~~~&~~~$\Gamma_{-1}$~~~&~~~$X_{-1}$~~~&~~~$Y_{-1}$~~~&~~~$M_{-1}$~~~\\
\hline\hline
A&1&1&0&0&0&0\\
&-1&0&1&1&1&1\\
\hline
B&1&1&0&1&0&1\\
&-1&0&1&0&1&0\\
\hline
C&1&1&0&1&1&0\\
&-1&0&1&0&0&1\\
\hline
D&1&1&0&0&1&1\\
&-1&0&1&1&0&0\\
\hline
\end{tabular}
\caption{Multiplicities of the rotation eigenvalues at  $\mathcal{C}_2$-invariant momenta for symmetry-adapted Wannier functions at Wyckoff position $W$ and with rotation eigenvalue $\lambda$.}\label{tbl:c2multi}
\end{table}
\subsection{Integer crystalline topological invariants in $\mathcal{C}_3$-symmetric crystals}
In a 2D $\mathcal{C}_3$-symmetric crystal we find three $\mathcal{C}_3$-symmetric Wyckoff positions per unit cell. These are labeled as $A, B, $ and $C$, , see Fig.~\ref{fig:wyckoff}(c). As before, we assume to have found symmetry-adapted Wannier functions, and let $N_{A;\lambda}$, $N_{B;\lambda}$, and$N_{C;\lambda}$ denote the total number of occupied symmetry-adapted Wannier functions at $A, B$, and $C$, with three-fold rotation eigenvalue $\lambda$. To ensure immunity against continuous deformations of the type shown in Fig.~1(b) in the main text, which correspond to $N_{W;\lambda}\rightarrow N_{W;\lambda}\pm 1\forall \lambda$, we are led to define a new set of six integers:
\begin{align*}
\nu_{W;\bar{r}}=-2N_{W;\bar{r}}+\sum_{r\neq\bar{r}}N_{W;r}\ \bar{r}=1,e^{i2\pi/3}~;\ W=A,B,C.
\end{align*}
As before, we can express these in terms of the multiplicities $\Gamma_{\bar{r}}$, $K^+_{\bar{r}}$, and $K^-_{\bar{r}}$ of the threefold rotation eigenvalues $\bar{r}$ at the high-symmetry points $\Gamma, K^+,$ and $K^-$. These relations read:
\begin{align*}
\nu_{A,1}&=-2\Gamma_1-\Gamma_{e^{i2\pi/3}}-\Gamma_{e^{i4\pi/3}}+K^+_{e^{i2\pi/3}}+K^+_{e^{i4\pi/3}}\\
&+K^-_{e^{i2\pi/3}}+K^-_{e^{i4\pi/3}},\\
\nu_{A,e^{i2\pi/3}}&=\Gamma_1+\Gamma_{e^{i4\pi/3}}-K^+_{e^{i2\pi/3}}-K^-_{e^{i2\pi/3}},\\
\nu_{B,1}&=\Gamma_{e^{i2\pi/3}}+\Gamma_{e^{i4\pi/3}}-K^+_{e^{i2\pi/3}}-K^-_{e^{i4\pi/3}},\\
\nu_{B,e^{i2\pi/3}}&=-\Gamma_{e^{i2\pi/3}}-K^+_{e^{i4\pi/3}}+K^-_{e^{i2\pi/3}}+K^-_{e^{i4\pi/3}},\\
\nu_{C,1}&=\Gamma_{e^{i2\pi/3}}+\Gamma_{e^{i4\pi/3}}-K^+_{e^{i4\pi/3}}-K^-_{e^{i2\pi/3}},\\
\nu_{C,e^{i2\pi/3}}&=-\Gamma_{e^{i2\pi/3}}+K^+_{e^{i2\pi/3}}+K^+_{e^{i4\pi/3}}-K^-_{e^{i4\pi/3}}.
\end{align*}
These formulas have been obtained by considering the multiplicities of the rotation eigenvalues for a single symmetry-adapted Wannier function centered at one of the Wyckoff positions, see Table~\ref{tbl:c3multi}.
\begin{table}[h!]
\begin{tabular}{|c|c||c|c|c|c|c|c|c|}
\hline
$W$&$\lambda$&$\Gamma_1$&$\Gamma_{e^{i\frac{2\pi}{3}}}$&$\Gamma_{e^{i\frac{2\pi}{3}}}$&$K^+_{e^{i\frac{2\pi}{3}}}$&$K^+_{e^{i\frac{2\pi}{3}}}$&$K^-_{e^{i\frac{2\pi}{3}}}$&$K^-_{e^{i\frac{2\pi}{3}}}$\\
\hline\hline
A&1&1&0&0&0&0&0&0\\
&$e^{i\frac{2\pi}{3}}$&0&1&0&1&0&1&0\\
&$e^{i\frac{4\pi}{3}}$&0&0&1&0&1&0&1\\
\hline
B&1&1&0&0&1&0&0&1\\
&$e^{i\frac{2\pi}{3}}$&0&1&0&0&1&0&0\\
&$e^{i\frac{4\pi}{3}}$&0&0&1&0&0&1&0\\
\hline
C&1&1&0&0&0&1&1&0\\
&$e^{i\frac{2\pi}{3}}$&0&1&0&0&0&0&1\\
&$e^{i\frac{4\pi}{3}}$&0&0&1&1&0&0&0\\
\hline
\end{tabular}
\caption{Multiplicities of the rotation eignevalues at $\mathcal{C}_3$-invariant momenta for symmetry-adapted Wannier functions at Wyckoff position $W$ and with rotation eigenvalue $\lambda$.}\label{tbl:c3multi}
\end{table}
\subsection{Integer crystalline topological invariants in $\mathcal{C}_6$-symmetric crystals}
In a $\mathcal{C}_6$-symmetric crystal, each unit cell hosts a single six-fold invariant Wyckoff position $A$, two three-fold invariant positions $B$ and $D$, and three two-fold invariant points $C, E,$ and $F$,  see Fig.~\ref{fig:wyckoff}(d). However, since all three- and two-fold positions are related by the six-fold symmetry, we only have two provide the number of symmetry-adapted Wannier functions at $A$, $B$, and $C$. As before, we require that the invariants are immune to transformations of the type shown in Fig.~1(b) in the main text. Hence, we find the following invariants:
\begin{align*}
\nu_{A;\bar{r}}&=-5N_{A;\bar{r}}+\sum_{r\neq\bar{r}}N_{A;r}\ \bar{r}=1,e^{i\pi/3},e^{i2\pi/3},-1,e^{i4\pi/3},\\
\nu_{B;\bar{r}}&=-2N_{B;\bar{r}}+\sum_{r\neq\bar{r}}N_{B;r}\ \bar{r}=1,e^{i2\pi/3},\\
\nu_{C;1}&=-N_{C;1}+N_{C;-1}.
\end{align*}
Expressed in terms of the multiplicities of the rotation eigenvalues at high-symmetry points, we find:
\begin{align*}
\nu_{A;1}&=-5\Gamma_{1}-\frac{5}{2}\Gamma_{e^{i\pi/3}}-\Gamma_{e^{i2\pi/3}}-\frac{1}{2}\Gamma_{-1}-\Gamma_{e^{i4\pi/3}}-\frac{5}{2}\Gamma_{e^{i5\pi/3}}+2K^+_{e^{i2\pi/3}}+2K^+_{e^{i4\pi/3}}+\frac{3}{2}M_{-1},\\
\nu_{A;e^{i\pi/3}}&=\Gamma_{1}-\frac{3}{2}\Gamma_{e^{i\pi/3}}+\Gamma_{e^{i2\pi/3}}+\frac{5}{2}\Gamma_{-1}+3\Gamma_{e^{i4\pi/3}}-2K^+_{e^{i2\pi/3}}-\frac{3}{2}M_{-1},\\
\nu_{A;e^{i2\pi/3}}&=\Gamma_{1}-\frac{1}{2}\Gamma_{e^{i\pi/3}}-3\Gamma_{e^{i2\pi/3}}-\frac{1}{2}\Gamma_{-1}+\Gamma_{e^{i4\pi/3}}-2K^+_{e^{i4\pi/3}}+\frac{3}{2}M_{-1},\\
\nu_{A;-1}&=\Gamma_{1}+\frac{1}{2}\Gamma_{e^{i\pi/3}}-\Gamma_{e^{i2\pi/3}}-\frac{7}{2}\Gamma_{-1}-\Gamma_{e^{i4\pi/3}}+2K^+_{e^{i2\pi/3}}+2K^+_{e^{i4\pi/3}}-\frac{3}{2}M_{-1},\\
\nu_{A;e^{i4\pi/3}}&=\Gamma_{1}+\frac{3}{2}\Gamma_{e^{i\pi/3}}+\Gamma_{e^{i2\pi/3}}-\frac{1}{2}\Gamma_{-1}-3\Gamma_{e^{i4\pi/3}}-2K^+_{e^{i2\pi/3}}+\frac{3}{2}M_{-1},\\
\nu_{B;1}&=\Gamma_{e^{i\pi/3}}+\Gamma_{e^{i2\pi/3}}+\Gamma_{e^{i4\pi/3}}-K^+_{e^{i2\pi/3}}-K^+_{e^{i4\pi/3}},\\
\nu_{B;e^{i2\pi/3}}&=-\Gamma_{e^{i\pi/3}}-\Gamma_{e^{i4\pi/3}}+K^+_{e^{i2\pi/3}},\\
\nu_{C;1}&=\frac{1}{2}\Gamma_{e^{i\pi/3}}+\frac{1}{2}\Gamma_{-1}-\frac{1}{2}M_{-1}.
\end{align*}
As before, these formulas have been obtained by considering the multiplicities of the rotation eigenvalues for single symmetry-adapted Wannier functions centered at one of the Wyckoff positions, see Table~\ref{tbl:c6multi}.
\begin{table}[t]
\begin{tabular}{|c|c||c|c|c|c|c|c|c|c|c|c|c|}
\hline
W&$\lambda$&$\Gamma_1$&$\Gamma_{e^{i\frac{\pi}{3}}}$&$\Gamma_{e^{i\frac{2\pi}{3}}}$&$\Gamma_{-1}$&$\Gamma_{e^{i\frac{4\pi}{3}}}$&$\Gamma_{e^{i\frac{5\pi}{3}}}$&$M_{-1}$&$K_{e^{i\frac{2\pi}{3}}}$&$K_{e^{i\frac{4\pi}{3}}}$\\
\hline\hline
A&1&1&0&0&0&0&0&0&0&0\\
&$e^{i\frac{\pi}{3}}$&0&1&0&0&0&0&1&1&0\\
&$e^{i\frac{2\pi}{3}}$&0&0&1&0&0&0&0&0&1\\
&-1&0&0&0&1&0&0&1&0&0\\
&$e^{i\frac{4\pi}{3}}$&0&0&0&0&1&0&0&1&0\\
&$e^{i\frac{5\pi}{3}}$&0&0&0&0&0&1&1&0&1\\
\hline
B+D&1&1&0&0&1&0&0&1&1&1\\
&$e^{i\frac{2\pi}{3}}$&0&1&0&0&1&0&1&0&1\\
&$e^{i\frac{4\pi}{3}}$&0&0&1&0&0&1&1&1&0\\
\hline
C+E+F&1&1&0&1&0&1&0&2&1&1\\
&-1&0&1&0&1&0&1&1&1&1\\
\hline
\end{tabular}
\caption{Multiplicities of the rotation eigenvalues at $C_6$-invariant momenta for symmetry-adapted Wannier functions at Wyckoff position $W$ and with rotation eigenvalue $\lambda$.}\label{tbl:c6multi}
\end{table}
\subsection{Integer crystalline topological invariants in $\mathcal{C}_n\mathcal{M}_z$-symmetric crystals, with $n=2,4,$ and $6$.}
Although, the analysis presented above applies to $\mathcal{C}_n$-symmetric crystals, however, essentially the same invariants characterize the symmetry-adapted Wannier functions in crystals that are $\mathcal{C}_n\mathcal{M}_z$-symmetric, with $n=2,4,$ and $6$ and $\mathcal{M}_z$ the out-of-plane reflection symmetry. 
\section{Corner charge of rotation-symmetric 2D crystals}
\begin{figure}[t]\centering
\includegraphics[width=.9\textwidth]{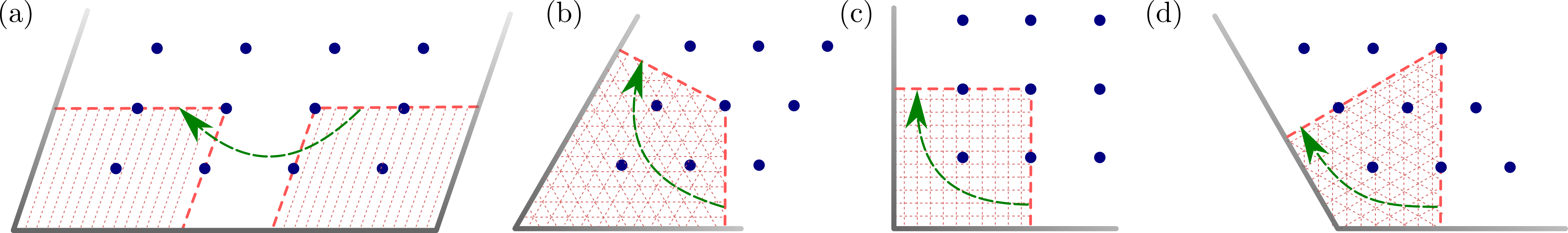}
\caption{Corners with edges related by $\mathcal{C}_m$-symmetry. The blue dots denote Wyckoff positions $A$. The charge inside the dashed region is then given by $\nu_A/m$. (a) corresponds to $m=2$, (b) $m=3$, (c) $m=4$, and (d) $m=6$.}
\label{fig:corner}
\end{figure}

For 1D inversion-symmetric insulators we have shown that the $\mathbb{Z}_2$ part of the invariants $\nu_A$ and $\nu_B$ can be related to the edge charges. Now suppose we consider  a 2D crystalline insulator with a $\mathcal{C}_n$-symmetric bulk. Let $\nu_{A}$ denote the $\mathbb{Z}_m$ part of an invariant associated with a Wyckoff position $A$. Now in general, if $A$ is invariant under an $m$-fold rotation, then the corner charge, measured with respect to $A$, is given by $\nu_A/m$. Indeed, we have explicitly shown this in the main text for a $\mathcal{C}_4$-symmetric crystal with $A$ a $\mathcal{C}_4$-invariant Wyckoff position. The same reasoning, applies to $\mathcal{C}_3$ and $\mathcal{C}_6$-symmetric corners. As stressed in the main text, it is crucial that the two edges that intersect at the corner are related by the same $m$-fold rotational symmetry. This is also illustrated schematically in Fig.~\ref{fig:corner}(b), (c), and (d).  Special care has to be taken if $m=2$, since two edges related by $\mathcal{C}_2$ do not intersect each-other. In this case, one should consider the sum of the lower-left corner charge and lower-right corner charge, both measured with respect to the same two-fold symmetric Wyckoff position $A$, see Fig.~\ref{fig:corner}(a).

\section{Corner charge pumps and chiral hinge states}
\begin{figure}[b]\centering
\includegraphics[width=.9\textwidth]{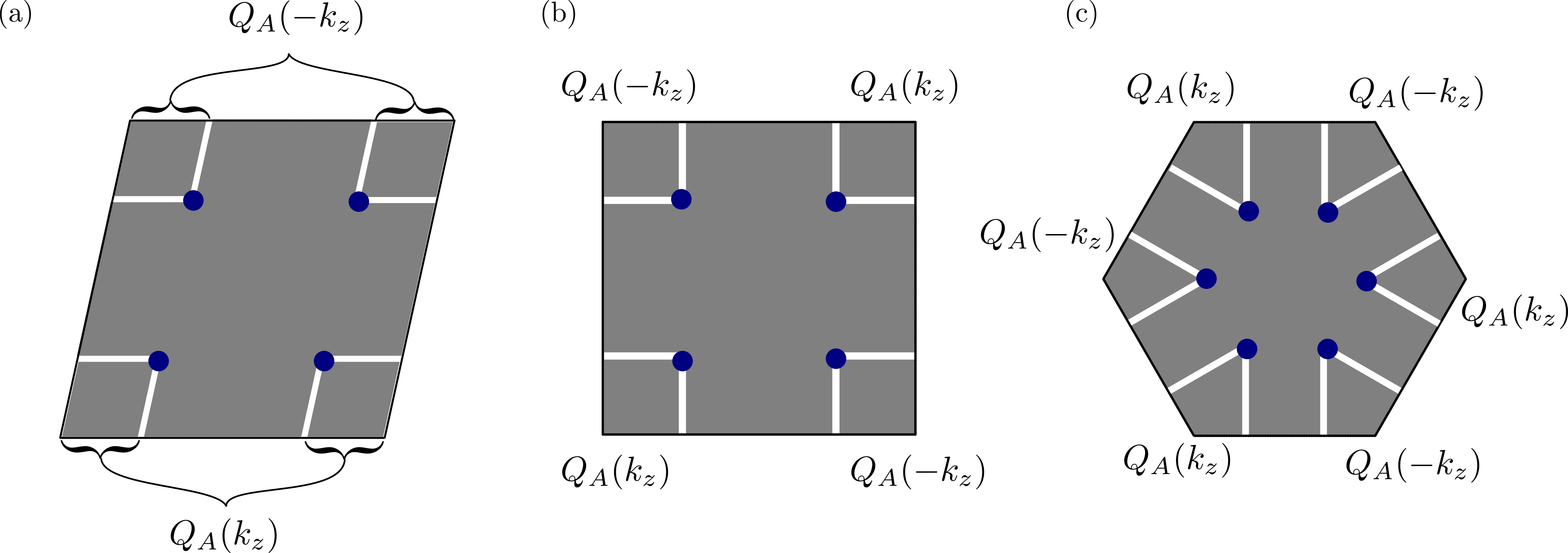}
\caption{Corner charge $Q_A(k_z)$ with respect to Wyckoff position $A$,  shown in blue, for a given $k_z$. (a), (b), and (c) correspond to $\mathcal{S}_2$, $\mathcal{S}_4$, and $\mathcal{S}_6$, respectively.}
\label{fig:hinge}
\end{figure}

Let us consider a 3D $\mathcal{S}_n=\mathcal{C}_n\mathcal{M}_z$-symmetric insulator, with $n=2,4,$ or $6$. Here, $\mathcal{S}_2$ is simply inversion, $\mathcal{S}_4$ a four-fold roto-inversion, while $\mathcal{S}_6$ corresponds to a three-fold roto-inversion, which therefore is both inversion and three-fold rotationally symmetric. Such a 3D crystal can be seen as a collection of $2$D systems with Hamiltonians $\mathcal{H}(k_z)$ parametrized by $k_z$. In particular, for $k_z=0,\pi$ we find that the 2D Hamiltonian is $\mathcal{S}_n$-symmetric. Now let us consider a geometry with surfaces that are related by the (roto)-inversion symmetry. Then, we find that the corner charge $Q_A(k_z)$, with $A$ a Wyckoff position invariant under $\mathcal{S}_n$, satisfies the following relation:
\begin{align*}
Q_A(k_z)=-Q_A(-k_z) \textrm{ mod }\dfrac{2}{n}.
\end{align*}
This formula follows from the following observations: \textit{(i)} The total charge in the edge and bulk parts is integer, which implies that the sum of all corner charges is integer. \textit{(ii)} The corner charges at two-neighboring corners at $+k_z$ and $-k_z$ are identical, see also Fig.~\ref{fig:hinge}. \textit{(iii)} The corner charges at corners related by $\mathcal{S}_n^2$ are identical (since this operation leaves $k_z$ untouched).

Now imagine that $Q_A(\pi)-Q_A(0)=1/2 (0)\textrm{ mod }1$. Then, each hinge sandwiched between two surfaces related by the $\mathcal{S}_n$-symmetry will host an odd (even) number of chiral states. Since $Q_A(0)$ and $Q_A(\pi)$ are given by $\nu_A(0)/n$ and $\nu_A(\pi)/n$, respectively, this translates into the statement that if and only if $\nu_A(\pi)-\nu_A(0)=n/2 \textrm{ mod }1$, then the system is a non-trivial higher order topological insulator. Note, that for an $\mathcal{S}_2$-symmetric,  insulator, two hinges are sandwiched between two surfaces related by the $\mathcal{S}_2$-symmetry. The invariant does not pinpoint on which of these two hinges the chiral state will reside: This is simply a microscopic detail of the specific surface. 

\begin{figure}[bh!]\centering
\includegraphics[width=.9\textwidth]{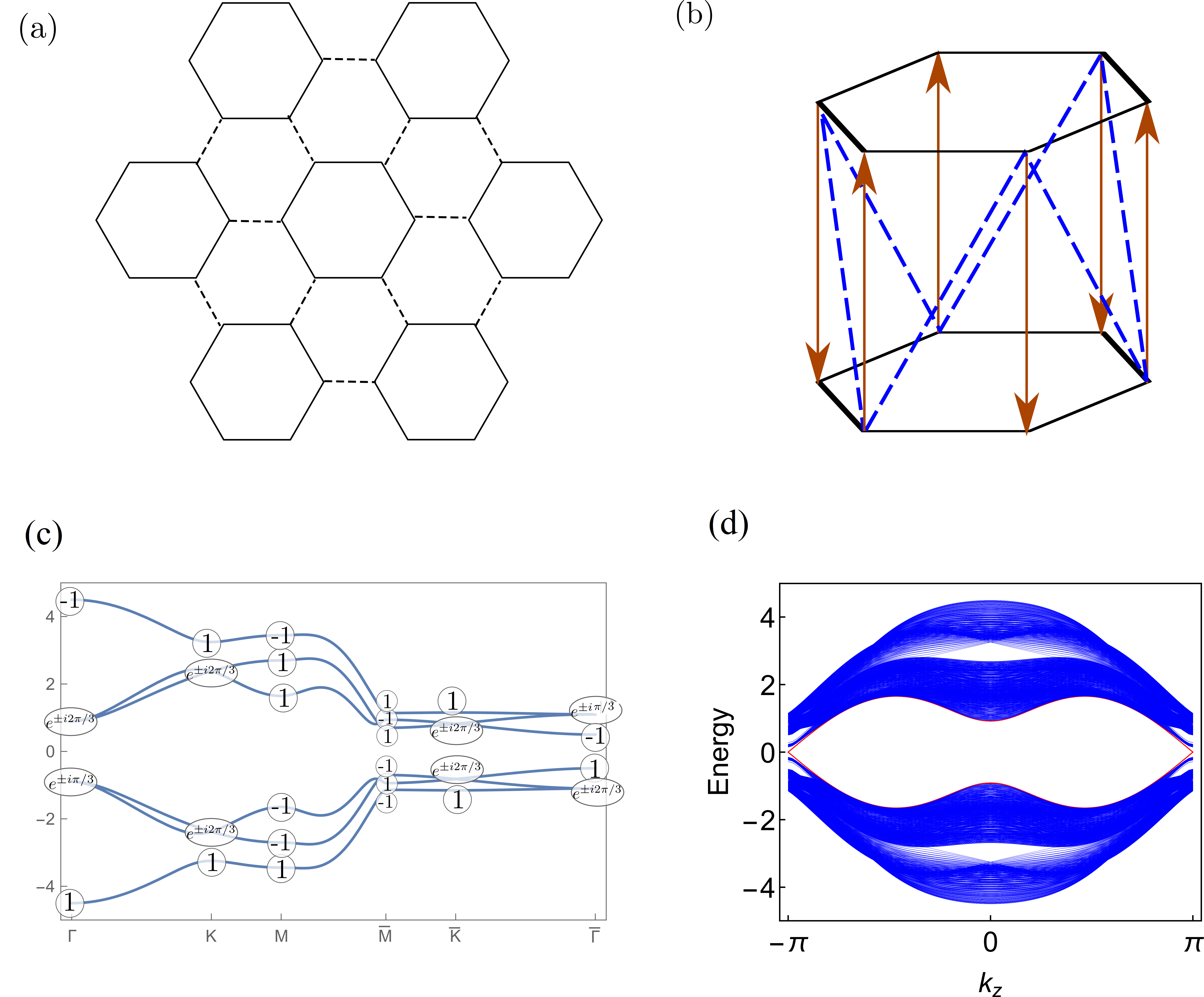}
\caption{(a) $\mathcal{C}_6$-symmetric Kekule distortion in graphene. Solid (dashed) lines correspond to intra (inter) unit cell hopping parameters $t_1$ ($t_2$). (b) Sketch of interlayer hopping parameters for a 3D higher-order topological insulator. Dashed lines corresponds to $t_4$, while the arrows correspond to $\pm i t_3$. Both break $\mathcal{C}_6$ and $\mathcal{M}_z$ individually, but respect their combination. The purely imaginary hopping parameters $\pm i t_3$ break time-reversal symmetry. (c) Bulk band structure along high-symmetry lines, with rotation eigenvalues at high-symmetry momenta. (d) Corresponding band structure for a system with periodic boundary conditions in the $\hat{z}$-direction, while open in the other directions. }
\label{fig:c6mz}
\end{figure}

Finally, we illustrate these finding for a $\mathcal{S}_6$-symmetric toy model. First, we consider the 2D honeycomb lattice with a Kekule distortion, such that the unit-cell hosts six sites, see Fig.~\ref{fig:c6mz}(a). The intra(inter)-unit cell hopping is given by $t_1$ ($t_2$). In the flat band limit, with $t_1\neq 0$ and $t_2=0$, we find that at half-filling the energies of the bands are given by $\pm 2 t_1, \pm t_1$,  with the latter doubly degenerate. Moreover, the topological invariant reads $\nu_A=3 \textrm{ mod }6$, and all others are zero. Similarly, we can consider the opposite limit, $t_1=0$, and $t_2\neq 0$. Then, the bands are triply degenerate, and have energies $\pm t_2$. Moreover, at half-filling we find $\nu_C=1 \textrm{ mod }2$, whereas all other invariants are zero. So in the latter case we expect that the corner charge measured with respect to $A$ vanishes, whereas in the former case, we expect that the corner charge is precisely equal to one half of the electron's charge. 

Next, let us consider a three-dimensional extension of this model. Here, we consider a 3D stack of the 2D lattices. We connect the stacks through purely imaginary nearest-neighbor hoppings $\pm i t_3$, as well as real diagonal next-nearest neighbor hoppings $t_4$, see Fig.~\ref{fig:c6mz}(b). These interlayer hopping break the $\mathcal{C}_6$ and $\mathcal{M}_z$ symmetries, but respect their combination $\mathcal{S}_6$. The Hamiltonian reads
\begin{align*}
\mathcal{H}_{\textbf{k}}&=-\begin{pmatrix}
2t_3 \sin(\textbf{k}\cdot\textbf{a}_3)&t_1+t_4 e^{i\textbf{k}\cdot\textbf{a}_3}&0&t_2 e^{i\textbf{k}\cdot(\textbf{a}_1-\textbf{a}_2)}&0&t_1+t_4 e^{i\textbf{k}\cdot\textbf{a}_3}\\
t_1+t_4 e^{-i\textbf{k}\cdot\textbf{a}_3}&-2t_3 \sin(\textbf{k}\cdot\textbf{b}_3)&t_1+t_4 e^{-i\textbf{k}\cdot\textbf{a}_3}&0&t_2 e^{i\textbf{k}\cdot\textbf{a}_1}&0\\
0&t_1+t_4 e^{i\textbf{k}\cdot\textbf{a}_3}&2t_3 \sin(\textbf{k}\cdot\textbf{a}_3)&t_1+t_4 e^{i\textbf{k}\cdot\textbf{a}_3}&0&t_2 e^{i\textbf{k}\cdot\textbf{a}_2}\\
t_2 e^{-i\textbf{k}\cdot(\textbf{a}_1-\textbf{a}_2)}&0&t_1+t_4 e^{-i\textbf{k}\cdot\textbf{a}_3}&-2t_3 \sin(\textbf{k}\cdot\textbf{a}_3)&t_1+t_4 e^{-i\textbf{k}\cdot\textbf{a}_3}&0\\
0&t_2 e^{-i\textbf{k}\cdot\textbf{a}_1}&0&t_1+t_4 e^{i\textbf{k}\cdot\textbf{a}_3}&2t_3 \sin(\textbf{k}\cdot\textbf{a}_3)&t_1+t_4 e^{i\textbf{k}\cdot\textbf{a}_3}\\
t_1+t_4 e^{-i\textbf{k}\cdot\textbf{a}_3}&0&t_2 e^{-i\textbf{k}\cdot\textbf{a}_2}&0&t_1+t_4 e^{-i\textbf{k}\cdot\textbf{a}_3}&-2t_3 \sin(\textbf{k}\cdot\textbf{a}_3)
\end{pmatrix}.
\end{align*}
In Fig.~\ref{fig:c6mz}(c) we plot the bulk band structure along HS lines, for $t_1=1$, $t_2=1.2$, $t_3=.8$, and $t_4=.9$. For these parameters, we find that in the $k_z=0$ plane, the invariant $\nu_A=3 \textrm{ mod }6$, and for $k_z=0$, we find $\nu_A=0 \textrm{ mod } 6$. Hence, we expect the presence of chiral hinge states for a geometry of the type shown in Fig.~\ref{fig:hinge}(c). This is indeed the case, see Fig.~\ref{fig:c6mz}(d), where we plot the corresponding band structure with periodic boundary conditions in the $z$ direction.


\begin{thebibliography}{10}
\bibitem{note1}
Since the Wyckoff positions $B$ and $D$ are transformed into each other by a fourfold rotation $N_{B; \pm 1} \equiv N_{D ; \pm1}$.
\bibitem{noteNANC}
The $\mathcal{S}_4$ symmetry ensures $N_A(k_z)+N_C(k_z)+2N_B(k_z)=N_F $ for $k_z=0,\pi$, where $N_W(k_z)=\sum_rN_{W;r}(k_z)$. It follows that $N_A(\pi)-N_A(0)=N_C(0)-N_C(\pi)+2[N_B(0)-N_B(\pi)]$. Since $\nu_B(\pi)=\nu_B(0)$, we find $N_B(\pi)=N_B(0) \textrm{ mod }2$. This in turn implies $\nu_{A,\bar{r}}(\pi)-\nu_{A,\bar{r}}(0)=\nu_{C,\bar{r}}(\pi)-\nu_{C,\bar{r}}(0) \textrm{ mod }4$, where we have used that $N_{A(C)}(k_z=0,\pi)=\nu_{A(C);\bar{r}}(k_z=0,\pi)\textrm{ mod }4$.
\end{thebibliography}
\end{document}